# Steady Floquet-Andreev States Probed by Tunnelling Spectroscopy

Sein Park[1,*], Wonjun Lee[1,*], Seong Jang[1], Yong-Bin Choi[1], Jinho Park[1], Woochan Jung[1], Kenji Watanabe[2], Takashi Taniguchi[3], Gil Young Cho[1,4,5,†] and Gil-Ho Lee[1,4,†]

[1] Department of Physics, Pohang University of Science and Technology, Pohang, Republic of Korea

[2] Research Center for Functional Materials, National Institute for Materials Science, Tsukuba, Japan

[3] International Center for Materials Nanoarchitectonics, National Institute for Materials Science, Tsukuba, Japan

[4] Asia Pacific Center for Theoretical Physics, Pohang, Republic of Korea

[5] Center for Artificial Low Dimensional Electronic Systems, Institute for Basic Science (IBS), Pohang 37673, Republic of Korea

[*] These authors contributed equally.

[†] Correspondence and requests for materials should be addressed to G.Y.C. (gilyoungcho@postech.ac.kr) or G.-H.L. (lghman@postech.ac.kr).

**Engineering quantum states through light-matter interaction has created a new paradigm in condensed matter physics. A representative example is the Floquet-Bloch state, which is generated by time-periodically driving the Bloch wavefunctions in crystals. Previous attempts to realise such states in condensed matter systems have been limited by the transient nature of the Floquet states produced by optical pulses, which masks the universal properties of non-equilibrium physics. Here, we report the generation of steady Floquet Andreev (F-A) states in graphene Josephson junctions by continuous microwave application and direct measurement of their spectra by superconducting tunnelling spectroscopy. We present quantitative analysis of the spectral characteristics of the F-A states while varying the phase difference of superconductors, temperature, microwave frequency and power. The oscillations of the F-A state spectrum with phase difference agreed with our theoretical calculations. Moreover, we confirmed the steady nature of the F-A states by establishing a sum rule of tunnelling conductance, and analysed the spectral density of Floquet states depending on Floquet interaction strength. This study provides a basis for understanding and engineering non-equilibrium quantum states in nano-devices.**

Light is a powerful tool for tailoring new quantum states by driving them out of equilibrium[1,2]. One prominent example is Floquet engineering, which involves dynamic control of the properties of quantum matter through a time-periodic electromagnetic drive. Similar to the space-periodic potential that copies Bloch energy bands along the crystal momentum direction, the time-periodic potential copies Floquet states along the energy direction. The Floquet state promises rapid and comprehensive control of the excitation spectrum and topological nature of physical systems by simply tuning the intensity, frequency and polarisation of the light. Therefore, there has been a great deal of research effort expended on the realisation and manipulation of a wide variety of long-sought-after quantum states, such as chiral topological orders with no equilibrium counterpart[3], including Floquet Majorana fermions[4] and a new braiding protocol in the energy dimension[5]. For example, the realisation and control of Floquet dynamics have been reported in both photonic and ultra-cold atomic systems[6, 7]. Another important class of platforms are solid-state systems, in which transient Floquet states have been mainly investigated. As Floquet interaction strength is proportional to the ratio of the electric field to the square of the frequency of light[8], previous condensed matter experiments realising Floquet states at the optical frequency of 30–50 THz have applied pulsed lasers to achieve a large electric field of $2.4–4.0 \times 10^7$ Vm$^{-1}$ (Refs.[9-11]). However, Floquet states persist no longer than an order of picoseconds due to the transient nature of pulsed laser and their inherently short lifetime. This short timescale makes it difficult to fully investigate and understand these novel light-driven states and exploit them for practical applications. Another critical issue affecting most experimental realisations of Floquet states is heating by the energy absorption from time-periodic driving. The driving increases the entropy density of the system and eventually brings it to a featureless state with no local correlations[3, 12]. Although a number of ways to reduce or avoid heating effects have been suggested[3, 13-18], the heating problem still remains in photonic and ultra-cold gas systems because they are well isolated from the outside environment. On the other hand, condensed matter systems have well-defined electron cooling paths, such as electron-phonon coupling and Wiedemann-Franz cooling of conducting electrons. However, a large electric field in an optical domain, which has been used in previous condensed matter experiments, requires pulsed measurements to avoid heating problems. One strategy to avoid thermal problems is to use lower frequency driving that requires a smaller electric field. There have been experimental studies in the microwave domain, but they relied on indirect methods for probing Floquet states either by tracing their time evolution[19, 20], qubit-resonator resonance conditions[21], magnetic resonance conditions[22] or AC Josephson effect[23-25].

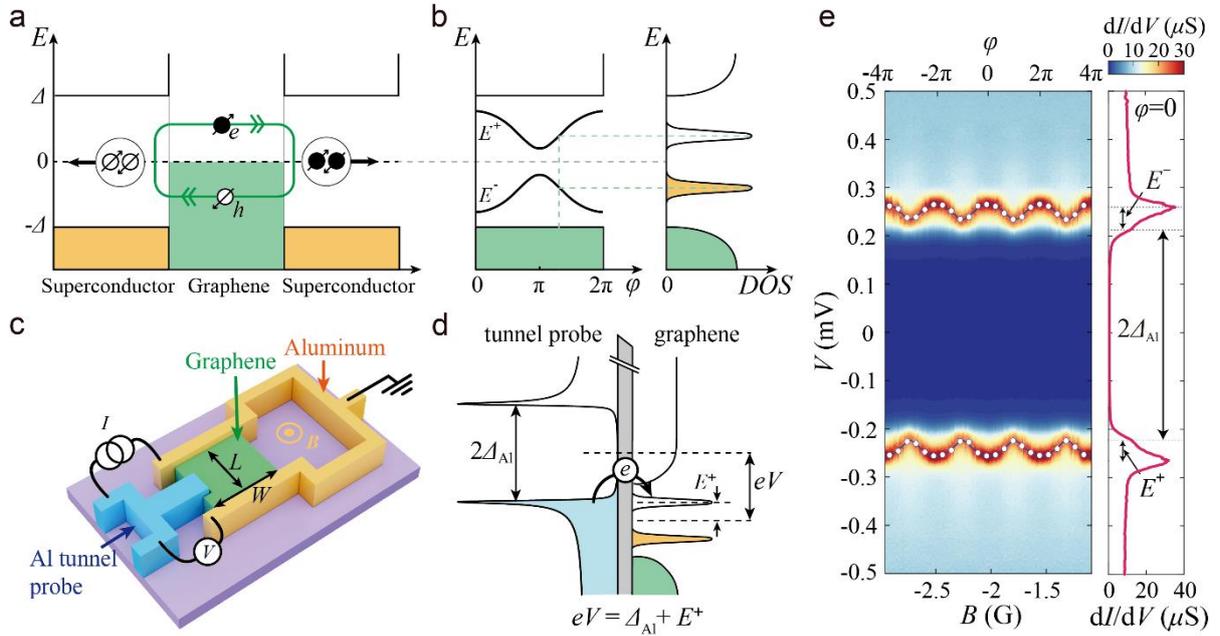

**Fig. 1 | Schematics of Andreev bound state and device geometry. a**, Microscopic illustration of an Andreev bound state (ABS) formed in graphene in contact with two superconductors. Within the superconducting gap $\Delta$, phase coherent electron-like (e) and hole-like (h) quasi-particles undergo Andreev reflection and subsequently form an ABS. **b**, In a short junction limit, a pair of upper ($E^+$) and lower ($E^-$) bands of ABS oscillates as a function of the phase difference of two superconductors, $\varphi$, as depicted in the left panel. The density of state (DOS) of ABS at a given $\varphi$ is shown in the right panel. **c**, Schematic of the device showing that the graphene (green) is in ohmic contact with aluminium (Al) superconducting electrodes (yellow)

that form a Josephson junction, and in tunnel contact with the other Al electrode (blue). The magnetic field ($B$) threading a superconducting loop controls $\varphi$. The graphene Josephson junction has length $L = 0.51$ μm and width $W = 3.0$ μm. **d,** Schematic showing the tunnelling process when bias voltage $V < 0$ is applied between the tunnelling probe and graphene, such that the occupied DOS peak of the tunnel probe matches the empty DOS peak ($E^+$) of ABS. **e,** Colour-coded plot of differential conductance d$I$/d$V$ as a function of $V$ and $B$ for device 1. The top horizontal axis shows the superconducting phase difference $\varphi$ corresponding to $B$. Circles and solid lines represent the d$I$/d$V$ peaks and corresponding theoretical fittings in a short junction limit, respectively. The line cut at $\varphi = 0$ plotted in the right panel shows d$I$/d$V$ peaks coming from upper ($E^+$) and lower ($E^-$) bands of ABS.

In contrast to previous studies, we report the experimental realisation of truly steady Floquet-Andreev (F-A) states based on Andreev bound states formed in a graphene Josephson junction (GJJ) and their spectra based on direct tunnelling spectroscopy. Here, we generated steady F-A states by continuously applying a monochromatic microwave drive and probed them by superconducting tunnelling spectroscopy with high energy resolution. We investigated the behaviour of the F-A states at various microwave frequencies and powers, and showed that their spectral features agreed well with our theoretical calculations. For example, we corroborated the steady nature of the F-A states by establishing a sum rule of the measured tunnelling conductance. This study clearly demonstrated steady Floquet states and will have a substantial impact on different areas of physics, including topological condensed matter, cold atoms in optical traps and nonequilibrium quantum statistical physics. Our technique, when combined with rapidly developing microwave technologies, can be extended to Floquet-based novel quantum device applications.

A GJJ consists of a non-superconducting graphene layer sandwiched between two superconductors, as shown in Fig. 1a. The GJJ allows the Josephson supercurrent by forming Andreev bound states (ABSs) of electron- and hole-like quasi-particles in the graphene, which are correlated by Andreev reflections at the graphene/superconductor interfaces. In the short junction limit where GJJ channel length $L$ is much shorter than the superconducting coherence length $\xi = \hbar v_F/\Delta_{\text{ABS}}$, a single pair of ABSs forms within the superconducting gap of ohmic-contacted Al electrodes $\Delta_{\text{ABS}}$ and oscillates with the macroscopic quantum phase difference between two superconductors $\varphi$ as $E^\pm(\varphi) = \Delta_{\text{ABS}}\sqrt{(1 - D\sin^2(\varphi/2))}$ [26-29]. Here, $\hbar$ is a reduced Planck's constant, $v_F = 10^6$ ms$^{-1}$ is the Fermi velocity of graphene, and $D$ is the contact transparency. The schematic in Fig. 1b shows the oscillation of ABS in a short junction limit and the corresponding density of state (DOS) at a fixed $\varphi$.

In this study, we performed tunnelling spectroscopy on ABS formed in graphene using Al superconducting tunnel contact with the graphene edge, as shown schematically in Fig. 1c (see Supplementary Fig. 1). Direct deposition of Al onto the graphene edge forms a sufficiently high potential barrier for the tunnelling probe due to the large inter-atomic distance between graphene and Al atoms[30]. In this structure, $\varphi = 2\pi\Phi/\Phi_0$ was controlled by the external magnetic flux $\Phi$ threading the superconducting ring in which GJJ was embedded. Here, $\Phi_0 = h/2e$ is the superconducting flux quantum with Planck's constant $h$ and electron charge $e < 0$.

We measured the voltage difference ($V$) between the Al tunnel probe and graphene while biasing the current ($I$). The output impedance of the current source (1 GΩ) was much larger than the largest tunnelling resistance (16 MΩ) measured in this experiment. We obtained the tunnelling differential conductance ($dI/dV$) as a function of $V$, which represents the convolution of DOS of the ABS in GJJ and that of a tunnel probe[28]. Figure 1d depicts the condition of the maximum $dI/dV$, where the DOS peak of a tunnel probe matches ABS at the bias voltage $V = (\Delta_{Al} + E^+)/e$, and where $\Delta_{Al}$ is the superconducting gap of the Al tunnel probe (see Supplementary Fig. 2). The sharp peak in DOS of the tunnel probe near $\Delta_{Al}$ allows a high energy resolution in tunnelling spectroscopy. Here, we estimated an upper bound of energy resolution (11.0 μeV) with the half-width at half-maximum of background subtracted $dI/dV$ (see Supplementary Fig. 3), which is three orders of magnitude better than that of time-resolved photoemission method[9, 10].

Figure 1e shows $dI/dV$ measured at 20 mK as a function of $V$ and $B$, the latter of which controls $\varphi(B) = 2\pi(B - B_0)A/\Phi_0$ with a magnetic field offset $B_0$ of a superconducting solenoid magnet and an effective area of the superconducting ring $A$. A clear oscillation of the $dI/dV$ peak as a function of $\varphi$ is shown, which fits well with our theoretical calculations (see Supplementary Fig. 4). In addition, the particle-hole symmetry of the superconductor manifests itself as a symmetric $dI/dV$ with respect to zero-bias voltage $V = 0$.

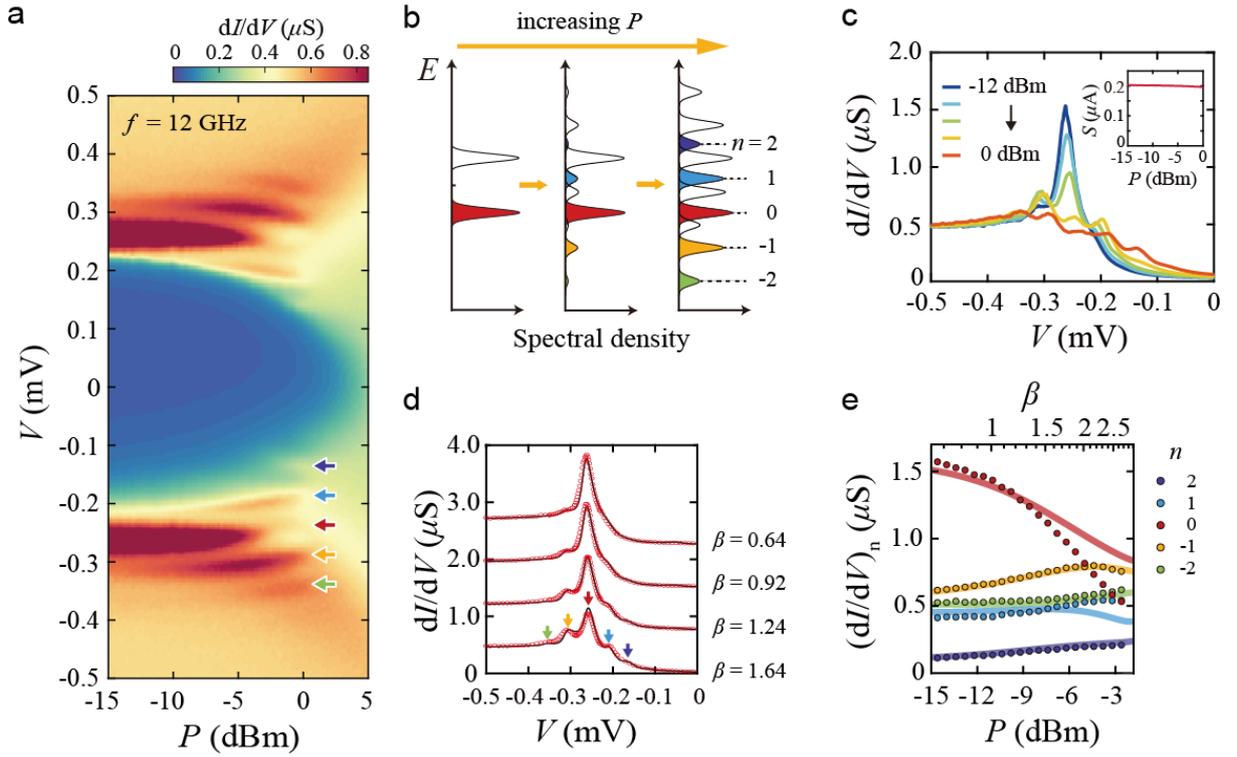

**Fig. 2 | Microwave power dependence of Floquet-Andreev states. a**, Colour-coded plot of differential conductance d$I$/d$V$ as a function of bias voltage $V$ and microwave power $P$, measured in device 2 with microwave frequency $f$ = 12 GHz. Floquet replicas of an Andreev bound state (ABS) indicated by arrows emerge as $P$ increases. **b**, Schematics of the evolution of the spectral density of Floquet-Andreev (F-A) states at different $P$. **c**, Line cuts of the plot in **a** at various $P$, from −12 to 0 dBm with 3-dBm intervals. Inset shows the integrated area of d$I$/d$V$ line cuts at a given $P$. **d** Line cuts of the plot in **a** at dimensionless Floquet interaction strength parameter $\beta$ = 0.64, 0.92, 1.24 and 1.64. Red circles are the experimental data and black lines are the theoretical fittings. Note that each line has an offset of 0.75 $\mu S$. **e**, d$I$/d$V$ values corresponding to the F-A states with the index $n = 0, \pm 1, \pm 2$ as a function of $P$. The colour of each line corresponds to the arrows in **a**.

To generate steady F-A states of GJJ, the devices were continuously irradiated with microwaves and the tunnelling spectrum of graphene was measured using a superconducting tunnel probe. As shown in Fig. 2a, additional peaks in d$I$/d$V$ emerged above and below the original ABS peaks as the microwave power $P$ increased. The average voltage spacing between the two adjacent d$I$/d$V$ peaks $\Delta V$ = 51.7 ± 11.0 μV at $P$ = -5.2 dBm corresponds to the microwave energy quanta $hf$ = 49.6 μeV with microwave frequency $f$ = 12 GHz. The error in $\Delta V$ is estimated by the half-width at half-maximum of background subtracted d$I$/d$V$. This behaviour can be understood by the emergence of Floquet replica states of the ABS, with the microwave irradiation acting as a time-periodic perturbation (Fig. 2b). As $P$ increased, the differential conductance of both the occupied lower band and unoccupied upper band of ABS were replicated to $E^{\pm} + nhf$ with integer $n = 0, \pm 1, \pm 2...$ In sharp contrast to previous experiments showing transient Floquet-Bloch states[9, 10], our system supports the steady F-A states generated by continuous Floquet driving, which are well resolved by tunnelling spectroscopy in DC measurements. The steadiness of F-A states is further evidenced by the sum rule of the tunnelling conductance: for a steady Floquet system, several non-trivial sum rules of physical observables have been proposed based on theoretical studies. We showed analytically that experimentally measured d$I$/d$V$ should satisfy a sum rule, *i.e.*, $S = \int_0^{\pm\infty}(dI/dV)\,dV$ is a constant if F-A states are steady (see Methods). The sum rule is based on the fact that the integral of the spectral weights of the F-A states over the energies is a constant[31]. Indeed, $S$ of our experimental data remained almost constant over a broad range of $P$ (see Fig. 2c) and various microwave frequencies (see Supplementary Fig. 5). The slight decrease of $S$ with increasing $P$ shown in the inset of Fig. 2c can be attributed to the DOS of high-energy F-A states leaking out from the finite integration window used for calculating $S$ [−0.5 mV, 0 mV].

Nevertheless, the overall agreement with the sum rule supports the steadiness of F-A states.

We proceeded to calculate the differential conductance d$I$/d$V$ theoretically, and compared the results with experimental data. Two major factors that determine tunnelling conductance are the spectral weight of F-A states and background differential conductance (d$I$/d$V$)$_{BG}$ from the proximity-induced superconductivity in graphene. First, the contribution of the F-A states can be obtained by solving time-dependent Schrodinger equations of the GJJ under a periodic, unpolarised electromagnetic field. The spectral weight of the $n$-th F-A replica follows the squared Bessel function averaged by polarisation angle $\theta$, $\bar{J}_n \equiv 2/\pi \int_0^{\pi/2} |J_n(\beta \cos \theta)|^2 d\theta$. Here, $J_n$ is the Bessel function of the first kind, $\beta = ev_F|E|/\hbar \omega^2$ is the dimensionless parameter that characterises the Floquet interaction strength, $|E| = \alpha \cdot 10^{P/(20 \text{ dBm})}$ is the electric field amplitude of the microwave, $\alpha$ describes the attenuation through the microwave input line and coupling efficiency between microwave antenna and the device and $\omega$ is the angular frequency of the microwave. Second, we identify (d$I$/d$V$)$_{BG}$, originating from the proximity effect of ohmic-contacted Al superconductors. To estimate (d$I$/d$V$)$_{BG}$ at different $P$, we note that microwave irradiation induces heating of the whole device structure. In the steady state, this energy inflow should be balanced with the cooling power by electron-phonon coupling, as the Wiedemann-Franz cooling to the superconducting electrodes is exponentially supressed due to their superconductivity and the radiation cooling is negligible at low temperature. Utilising the power-law relation between $P$ and $T$ for electron-phonon cooling in graphene[32-34] (see Supplementary Fig. 6), we estimated (d$I$/d$V$)$_{BG}$ at a given power $P$ from the d$I$/d$V$ at the corresponding temperature $T$, in the absence of microwave driving (see Supplementary Fig. 7). By summing these factors, we calculated d$I$/d$V$ for comparison with the experimental data, as shown in Fig. 2d (see Supplementary Section 7 for the detailed fitting procedure). Our theoretical calculation explains well both the overall shapes of d$I$/d$V$ and the peak positions and heights of F-A replicas for the broad range of $\beta$. The asymmetricity in the experimental data around zero-voltage bias is presumably due to the asymmetric tunnelling barrier. Therefore, we chose to fit the data only for the negative-voltage bias, where features from F-A replicas are better resolved. To quantitatively discuss the behaviour of F-A replicas with different Floquet interaction strengths, the experimental values of d$I$/d$V$ for the $n$-th F-A replica [(d$I$/d$V$)$_n$] are plotted as a function of $P$ in Fig. 2e, and feature non-monotonic dependence on the power $P$. This is well captured by our theoretical calculations up to about −7 dBm. Our theoretical modelling of the F-A states and d$I$/d$V$ is largely based on the static solutions for ABS, and assumes local equilibration of electrons (see Supplementary Section 7). The experimental data gradually deviates from the theoretical fitting at $P > -7$ dBm ($\beta \approx 1.7$); this can be attributed to, for instance, the significant heating of electrons that leads to a non-Fermi-Dirac distribution of electrons.

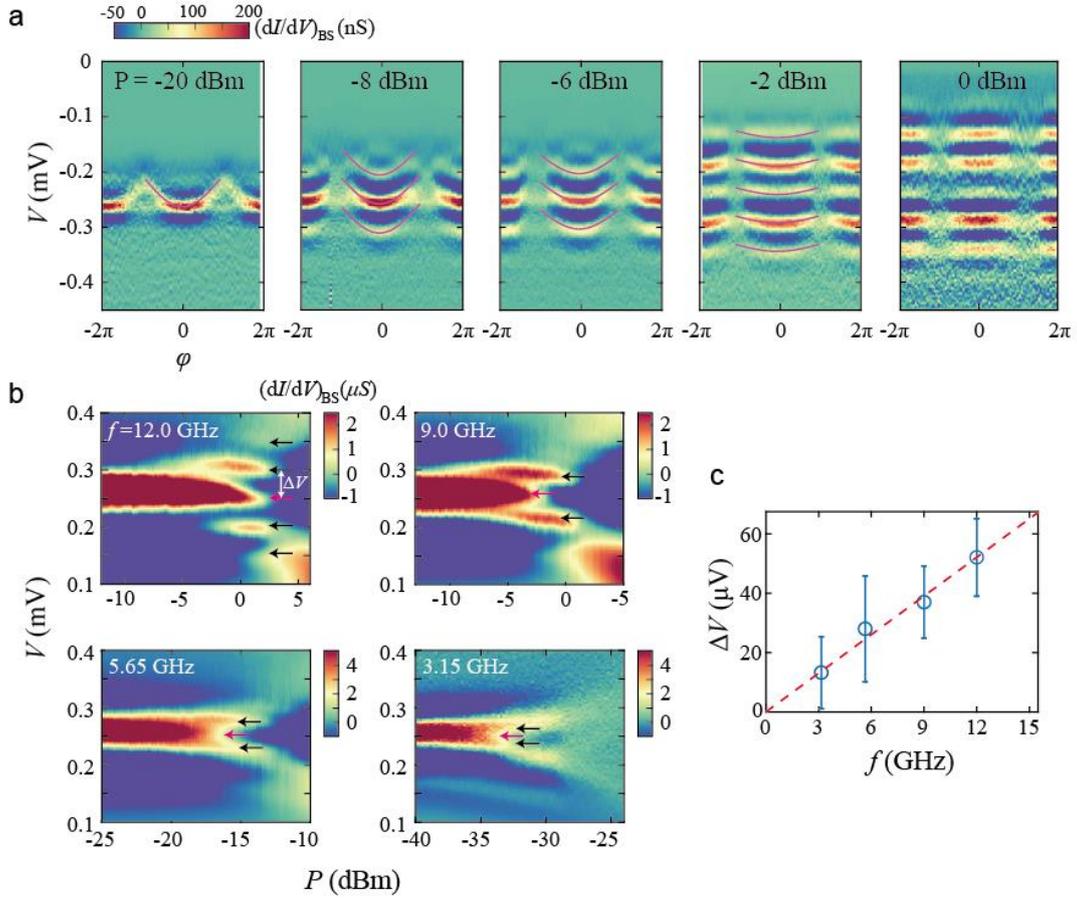

**Fig. 3 | Phase and frequency dependence of Floquet-Bloch states**. **a,** Background subtracted differential conductance $(dI/dV)_{BS}$ measured in device 2 as a function of bias voltage $V$ and phase difference $\varphi$, with various microwave power $P$. The microwave frequency is $f = 12$ GHz. For $P \geq -8$ dBm, $V$ is slightly shifted for easier comparison of the data. The theoretical calculation is overlaid as red solid lines. The colour scale ranges of $-2$ and $0$ dBm are changed to [$-30$ nS, $100$ nS] and [$-10$ nS, $40$ nS], respectively. **b,** $(dI/dV)_{BS}$ measured in device 1 as a function of $V$ and $P$ at various $f$. Original Andreev bound state (ABS) peak and its Floquet replicas are denoted by red and black arrows, respectively. **c,** The average voltage difference between adjacent peaks ($\Delta V$) in **b** is plotted as a function $f$. The errors are estimated by the half-width at half-maximum of background subtracted d$I$/d$V$. The red dotted line is a linear fit

crossing the origin.

We now discuss the superconducting phase dependence of F-A states. Figure 3a shows synchronised oscillations of the original ABS and Floquet replica states with phase difference $\varphi$, which further confirms that the F-A states are replications of the original ABS. For better visualisation of the oscillations, we subtracted the background obtained by averaging d$I$/d$V$ over several peaks in $V$ (see Supplementary Fig. 8). With increasing $P$, the peak values of the background subtracted differential conductance $(dI/dV)_{BS}$ of Floquet replica ($n \neq 0$) states became larger, while that of the original ABS ($n = 0$) became smaller (Fig. 2e). In addition, the oscillation amplitude, which corresponds to $\Delta_{ABS}$, becomes smaller with increasing $P$ and vanished at $P = 0$ dBm. This can be explained by the significant electron heating due to strong microwave irradiation, such that electron temperature reaches the critical temperature of the ohmic-contacted Al superconductor (see Supplementary Fig. 6). Solutions of the time-dependent Schrodinger equation of the GJJ under the unpolarised electromagnetic field allowed us to compute the quasi-energy spectrum of the F-A states. We found that our theoretical calculations well reproduced the observed oscillation of the F-A spectrum, as shown in Fig. 3a. Here, d$I$/d$V$ for $V < 0$ represents only the replicas of $E^+(\varphi)$, as electrons in the superconducting tunnel probe can hop only to replicas of $E^+(\varphi)$ that are empty (see Supplementary Fig. 9).

Finally, we discuss the microwave frequency dependence of F-A states. The power dependence of $(dI/dV)_{BS}$ at various microwave frequencies is shown in Fig. 3b. For all frequencies, the decreasing intensity of the peak from the original ABS ($n = 0$, red arrow) with increasing $P$ is accompanied by increasing intensity of the Floquet replica states ($n \neq 0$, white arrows), as expected from the sum rule and Bessel function behaviour. The overall peak position shift to lower voltages with increasing $P$ is attributed to the heating of electrons. We found that the voltage spacing $\Delta V$ between adjacent $(dI/dV)_{BS}$ peaks was linearly proportional to $f$ (Fig. 3c) with a slope of $4.34 \pm 0.46$ μV/GHz. This value is close to the slope expected for F-A states, $h/e = 4.14$ μV/GHz, confirming the nature of F-A states.

In sum, we have realised the steady F-A states in a Josephson junction device by continuous microwave irradiation without significant heating, and directly measured their energy spectra by superconducting tunnelling spectroscopy. This technique is readily applicable to other low-dimensional topological materials, such as topological insulators and topological semi-metals, for studying and engineering topological Floquet physics[5]. The manifestation of the topology of such Floquet states will be most evident in stationary, non-heated Floquet systems, which previous technologies in the optical domain could not achieve. Realisation of more exotic spatiotemporal-driven quantum states by circularly polarised or quasi-periodic driving, which have been pursued only in theory[35-38], is now experimentally possible. Our observations open up new avenues to design, build, measure and exploit the exotic far-from-equilibrium quantum states, and should lead to novel electronic device applications for microwave engineering of condensed matter.

**Methods**

**Device fabrication.** Graphene and hexagonal boron nitride (hBN) flakes were exfoliated on separate silicon oxide wafers. A graphene monolayer was encapsulated by hBN flakes using a dry transfer method[39]. The top hBN was 48 nm and the bottom hBN was 52 nm. The graphene stack was shaped by reactive ion etching with CF$_4$ and O$_2$ plasmas, and the electrodes were patterned by electron-beam lithography and deposited by electron beam evaporation onto freshly etched edges of graphene. Ohmic-contact electrodes consisted of a 5-nm Ti adhesion layer and 70-nm Al superconducting layer, the latter of which was deposited with BN-TiB$_2$ crucibles at a rate of 1.5 Ås$^{-1}$. A superconducting tunnel probe electrode consisting of a 70-nm Al layer without any adhesion layers was deposited with BN-TiB$_2$ crucibles at a rate of 0.3 Ås$^{-1}$. The evaporation chamber pressure during the evaporation was kept at less than $1 \times 10^{-7}$ mTorr.

**Theoretical calculation of differential conductance.** We theoretically calculated the time-averaged differential conductance of the ABS. We first modelled the GJJ without any external driving and obtained the ABS. We then included microwave driving within the graphene region and obtained the Floquet states for a given polarisation of the light. Next, we computed the spectral function and differential conductance of the Floquet states using the Floquet-Kubo formula.[3] Finally, we averaged the differential conductance over the polarisation for comparison with our experiments where the microwaves were unpolarised. For further details, please see Supplementary Fig. 7 and Supplementary Section 9.

We first calculated the ABS spectra and corresponding wavefunctions using standard methods.[40, 41] Essentially, the model is equivalent to "a (Dirac) particle in a box" under the superconducting boundary conditions. Formally, the Hamiltonian of graphene with superconducting electrodes is given as the Dirac-Bogoliubov-de Gennes (DBdG) equation.[42] To reflect the finite width of the GJJ, we allow several conduction modes to appear in the spectrum of ABS. Each mode is identified with a given y-directional momentum, i.e. the momentum along the width of the GJJ. We found that the 12 lowest modes are sufficient to explain the experiment, which gives small y-directional momentums compared to x-directional momentums. We numerically obtained the wavefunctions $|n, \varphi\rangle$ and energy spectrum of the ABS in a GJJ, $E_n(\varphi)$, where $\varphi$ is the phase difference between the two superconductors and $n$ is the index of the ABS. The resulting spectra are fairly independent of the graphene edge boundary conditions, so we present only the results with the metallic armchair boundary conditions along the y-direction. Next, we obtain the Floquet states of the GJJ in the presence of the electromagnetic gauge field $\vec{A}(t)$ in the graphene region. We obtain the Floquet states under the field by directly solving the time-dependent Schrödinger's equation. We then calculate the time-averaged spectral function of the Floquet states.[28, 31] The time-averaged spectral function $B_l(\omega)$ of a Floquet state with quasi-energy $\xi_l$ is given as

$$B_l(\omega) = \sum_{m=-\infty}^{\infty} |J_m(\beta)|^2 \, \delta\big(\omega - (\xi_l/\hbar + m\Omega)\big),$$

where $J_n(\beta)$ is the Bessel function of the first kind, $\beta$ is the dimensionless parameter $\beta \equiv \frac{evA_x}{\hbar\Omega}$ and $\Omega$ is the frequency of $\vec{A}(t)$.

Finally, we compute the differential conductance of the Floquet states. Here, our starting point is the standard expression for electronic tunnelling current in terms of Green's functions for electrons in the graphene and superconducting tunnel probe:[28]

$$I = -W \, \Re \int_{-\infty}^{\infty} d t' \Theta(t - t') e^{-i\frac{eV}{\hbar}(t-t')} \sum_{l,r} [\, G_{ll}^{>}(t',t) G_{rr}^{<}(t,t') - G_{ll}^{<}(t',t) G_{rr}^{>}(t,t')\,].$$

where $W = \frac{2e}{\hbar}|t_{\text{probe}}|^2$ and $t_{\text{probe}}$ is the tunnel matrix element, which is assumed to be a constant, subscript $l$ corresponds to the state of the superconducting tunnel probe and subscript $r$ corresponds to the label of a state of the graphene. The Green functions $G^{>,<}{}_{ll}(t',t)$ can be rewritten in terms of the spectral function of the superconducting tunnel probe[43] $A_T(\epsilon) \propto \Re[\frac{|\epsilon+i\gamma\Delta|}{\sqrt{(\epsilon+i\gamma\Delta)^2 - \Delta^2}}]$ and the Fermi-Dirac distribution function $f_{FD}(\epsilon)$. Here, $\gamma$ is the Dynes depairing parameter and $\Delta$ is the pairing gap. With this, the time-averaged tunnelling current at low temperature is given as

$$\langle I(V) \rangle_\tau = \frac{W}{\pi} \int_{-\infty}^{\infty} d\omega \sum_r [g(\xi_r) - \Theta(-\omega - \frac{eV}{\hbar})] A_{\text{probe}}(\omega + \frac{eV}{\hbar}) B_r(\omega).$$

Thus, the time-averaged differential conductance is given as

$$\left\langle \frac{dI}{dV} \right\rangle_\tau = \frac{e}{\pi\hbar} W \int_{-\infty}^{\infty} d\omega \sum_r [g(\xi_r) - \Theta(-\omega)] \partial_\omega A_T(\omega) B_r(\omega - \frac{eV}{\hbar}).$$

**Derivation of the sum rule of differential conductance.** Here, we describe our derivation of a sum rule for differential conductance, by following methods in the literature[3, 28, 31] that discuss closely related Floquet sum rules. The detailed derivation of the sum rule is given in the section 9 of the Supplementary Information.

The sum of the time-averaged differential conductance can be computed as

$$S \equiv \int_{-\infty}^{\infty} dV \left\langle \frac{dI}{dV} \right\rangle_\tau = \lim_{V \to \infty} [\langle I(V) \rangle_\tau - \langle I(-V) \rangle_\tau],$$

This can be computed using the tunnelling current obtained above, as

$$S = \frac{2e}{\pi\hbar}|t_\text{probe}|^2 \int_{-\infty}^{\infty} d\omega \sum_r B_r(\omega).$$

Note that $\int_{-\infty}^{\infty} d\omega \sum_r B_r(\omega)$ is the sum of the density of states of the Floquet states; it is known to respect a sum rule[31] and $S$ should therefore satisfy a sum rule (see Supplementary Section 9-3 for details). Finally, invoking the particle-hole symmetry of the overall spectrum, we conclude that $\int_0^\infty dV \left\langle \frac{dI}{dV} \right\rangle_\tau = \frac{1}{2}S$ also satisfies the sum rule, which was confirmed in our experimental data.

**Data availability**

The data supporting the findings of this study are available from the corresponding author upon reasonable request.

**Acknowledgements**

We thank H.-J. Lee, K.W. Kim, J.C.W. Song and C. Lee for critical reading of the manuscript. S.P., S.J., Y.-B.C. and G.-H.L. acknowledge the support of the Samsung Science and Technology Foundation (project no. SSTF-BA1702-05) for device fabrications and low-temperature measurements. W.J. and G.-H.L. acknowledge the support of National Research Foundation of Korea (NRF) funded by the Korean Government (Grant No. 2016R1A5A1008184, 2020R1C1C1013241 and 2020M3H3A1100839) for data analysis. W.L. and G.Y.C. acknowledge the support of the National Research Foundation of Korea (NRF) funded by the Korean Government (Grant No. 2020R1C1C1006048 and 2020R1A4A3079707), as well as Grant No. IBS-R014-D1. G.Y.C. is partially supported by the Air Force Office of Scientific Research under Award No. FA2386-20-1-4029. K.W. and T.T. acknowledge support from the Elemental Strategy Initiative conducted by the MEXT, Japan, Grant Number JPMXP0112101001 and JSPS KAKENHI Grant Number JP20H00354.

**Competing interests**

The authors declare no competing interests.

**Author Contributions**

G.-H.L. and G.Y.C. conceived and supervised the project. S.P. designed and fabricated the devices. T.T. and K.W. provided the hBN crystal. S.P., S.J. and Y.-B.C. performed the measurements. W.L. and G.Y.C. carried out theoretical calculations. S.P., W.L., S.J., J.P., W.J., G.Y.C. and G.-H.L. performed the data analysis. S.P., W.L., S.J., G.Y.C. and G.-H.L. wrote the paper.

**References**

1. Fausti, D., *et al.* Light-induced superconductivity in a stripe-ordered cuprate. *Science* **331,** 189 (2011).
2. Matsunaga, R., *et al.* Light-induced collective pseudospin precession resonating with Higgs mode in a superconductor. *Science* **345,** 1145 (2014).
3. Rudner, M. S. and Lindner, N. H. Band structure engineering and non-equilibrium

dynamics in Floquet topological insulators. *Nat Rev Phys* **2,** 229-244 (2020).

4. Jiang, L.*, et al.* Majorana Fermions in equilibrium and in driven cold-atom quantum wires. *Phys. Rev. Lett.* **106,** 220402 (2011).

5. Bauer, B.*, et al.* Topologically protected braiding in a single wire using Floquet Majorana modes. *Phys. Rev. B* **100,** 041102(R) (2019).

6. Clark, L. W.*, et al.* Interacting Floquet polaritons. *Nature* **571,** 532-536 (2019).

7. Wintersperger, K.*, et al.* Realization of an anomalous Floquet topological system with ultracold atoms. *Nat. Phys.* **16,** 1058-1063 (2020).

8. Freericks, J. K., Krishnamurthy, H. R. and Pruschke, T. Theoretical description of time-resolved photoemission spectroscopy: application to pump-probe experiments. *Phys. Rev. lett.* **102,** 136401 (2009).

9. Wang, Y. H., Steinberg, H., Jarillo-Herrero, P. and Gedik, N. Observation of Floquet-Bloch States on the surface of a topological insulator. *Science* **342,** 453 (2013).

10. Mahmood, F.*, et al.* Selective scattering between Floquet–Bloch and Volkov states in a topological insulator. *Nat. Phys.* **12,** 306-310 (2016).

11. Mciver, J. W.*, et al.* Light-induced anomalous Hall effect in graphene. *Nat. Phys.* **16,** 38-41 (2020).

12. D'alessio, L. and Rigol, M. Long-time behavior of isolated periodically driven interacting lattice systems. *Phys. Rev. X* **4,** 041048 (2014).

13. Abanin, D. A., De Roeck, W., Ho, W. W. and Huveneers, F. Effective Hamiltonians, prethermalization, and slow energy absorption in periodically driven many-body systems. *Phys. Rev. B* **95,** 014112 (2017).

14. Rubio-Abadal, A.*, et al.* Floquet prethermalization in a Bose-Hubbard system. *Phys.*

*Rev. X* **10,** 021044 (2020).

15. Mori, T., Ikeda, T. N., Kaminishi, E. and Ueda, M. Thermalization and prethermalization in isolated quantum systems: a theoretical overview. *J. Phys. B: At. Mol. Opt. Phys.* **51,** 112001 (2018).

16. Lazarides, A., Das, A. and Moessner, R. Equilibrium states of generic quantum systems subject to periodic driving. *Phys. Rev. E* **90,** 012110 (2014).

17. Ponte, P., Chandran, A., Papić, Z. and Abanin, D. A. Periodically driven ergodic and many-body localized quantum systems. *Annals of Physics* **353,** 196-204 (2015).

18. Nathan, F., Abanin, D., Berg, E., Lindner, N. H. and Rudner, M. S. Anomalous Floquet insulators. *Phys. Rev. B* **99,** 195133 (2019).

19. Deng, C., Orgiazzi, J.-L., Shen, F., Ashhab, S. and Lupascu, A. Observation of Floquet states in a strongly driven artificial atom. *Phys. Rev. Lett.* **115,** 133601 (2015).

20. Fuchs, G. D., Dobrovitski, V. V., Toyli, D. M., Heremans, F. J. and Awschalom, D. D. Gigahertz dynamics of a strongly driven single quantum spin. *Science* **326,** 1520 (2009).

21. Koski, J. V.*, et al.* Floquet spectroscopy of a strongly driven quantum dot charge qubit with a microwave resonator. *Phys. Rev. Lett.* **121,** (2018).

22. Jamali, S.*, et al.* Floquet spin states in OLEDs. *Nat. Commun.* **12,** 465 (2021).

23. Huang, K.-F.*, et al.* Interference of Cooper quartet Andreev bound states in a multi-terminal graphene-based Josephson junction. Preprint at https://arXiv.org/abs/2008.03419 (2020).

24. Melin, R., Danneau, R., Yang, K., Caputo, J. G. and Doucot, B. Engineering the Floquet spectrum of superconducting multiterminal quantum dots. *Phys. Rev. B* **100,** (2019).

25. Melin, R., Caputo, J. G., Yang, K. and Doucot, B. Simple Floquet-Wannier-Stark-

Andreev viewpoint and emergence of low-energy scales in a voltage-biased three-terminal Josephson junction. *Phys. Rev. B* **95,** (2017).

26. Nichele, F.*, et al.* Relating Andreev bound states and supercurrents in hybrid Josephson junctions. *Phys. Rev. lett.* **124,** 226801 (2020).

27. Bretheau, L.*, et al.* Tunnelling spectroscopy of Andreev states in graphene. *Nat. Phys.* **13,** 756-760 (2017).

28. Pillet, J. D.*, et al.* Andreev bound states in supercurrent-carrying carbon nanotubes revealed. *Nat. Phys.* **6,** 965-969 (2010).

29. Giazotto, F., Peltonen, J. T., Meschke, M. and Pekola, J. P. Superconducting quantum interference proximity transistor. *Nat. Phys.* **6,** 254-259 (2010).

30. Lee, G.-H., Kim, S., Jhi, S. H. and Lee, H.-J. Ultimately short ballistic vertical graphene Josephson junctions. *Nat. Commun.* **6,** 6181 (2015).

31. Uhrig, G. S., Kalthoff, M. H. and Freericks, J. K. Positivity of the spectral densities of retarded Floquet Green functions. *Phys. Rev. lett.* **122,** 130604 (2019).

32. Viljas, J. K. and Heikkilä, T. T. Electron-phonon heat transfer in monolayer and bilayer graphene. *Phys. Rev. B* **81,** 245404 (2010).

33. Chen, W. and Clerk, A. A. Electron-phonon mediated heat flow in disordered graphene. *Phys. Rev. B* **86,** 125443 (2012).

34. Walsh, E. D.*, et al.* Graphene-based Josephson-junction single-photon detector. *Phys. Rev. Applied* **8,** 024022 (2017).

35. Verdeny, A., Puig, J. and Mintert, F. Quasi-periodically driven quantum systems. *Zeitschrift für Naturforschung A* **71,** 897-907 (2016).

36. Crowley, P. J. D., Martin, I. and Chandran, A. Topological classification of

quasiperiodically driven quantum systems. *Phys. Rev. B* **99,** 064306 (2019).

37. Crowley, P. J. D., Martin, I. and Chandran, A. Half-integer quantized topological response in quasiperiodically driven quantum systems. *Phys. Rev. lett.* **125,** 100601 (2020).

38. Sentef, M. A., *et al.* Theory of Floquet band formation and local pseudospin textures in pump-probe photoemission of graphene. *Nat. Commun.* **6,** 7047 (2015).

39. Wang, L., *et al.* One-dimensional electrical contact to a two-dimensional material. *Science* **342,** 614 (2013).

40. Beenakker, C. W. J. Colloquium: Andreev reflection and Klein tunneling in graphene. *Rev. Mod. Phys.* **80,** 1337-1354 (2008).

41. Titov, M. and Beenakker, C. W. J. Josephson effect in ballistic graphene. *Phys. Rev. B* **74,** 041401(R) (2006).

42. Beenakker, C. W. Specular Andreev reflection in graphene. *Phys. Rev. lett.* **97,** 067007 (2006).

43. Dynes, R. C., Narayanamurti, V. and Garno, J. P. Direct measurement of quasiparticle-lifetime broadening in a strong-coupled superconductor. *Phys. Rev. Lett.* **41,** 1509-1512 (1978).